\documentclass{optica-article}

\journal{opticajournal} 

\articletype{Research Article}

\usepackage{lineno}
\usepackage{amsmath,mathtools}

\hypersetup{
    colorlinks=true,
    linkcolor=blue,
    citecolor=black,
    urlcolor=blue
}

\DeclareMathOperator*{\argmax}{arg\,max}

\begin{document}

\title{Valley-peak modulation in phase space: a law-invariant VPM and its theta-function structure}

\author{Aaron J. Hendrickson\authormark{1,*} and David P. Haefner\authormark{2}}

\address{\authormark{1}U.S. Navy, NAWCAD DAiTA Group, Atlantic Ranges \& Targets, Optical Systems, Electro-Optical Tracking Systems, 23013 Cedar Point Road, Bldg. 2118, Patuxent River MD, 20670\\
\authormark{2}U.S. Army Combat Capabilities Development Command (DEVCOM), C5ISR Center, Research \& Technology Integration (RTI), 10221 Burbeck Road, Fort Belvoir, VA 22060}

\email{\authormark{*}aaron.j.hendrickson2.civ@us.navy.mil}

\begin{abstract*}
Valley-peak modulation (VPM) was introduced as a metric for quantifying read noise in deep sub-electron read noise (DSERN) CMOS image sensors. In its original amplitude-domain definition VPM depends on both read noise and quanta exposure, yet Starkey and Fossum demonstrated exposure-independent approximations valid in the DSERN regime. Here we identify the invariant object those approximations probe, and find its invariance extends beyond exposure to the electron-number law itself. A phase mapping quotients the sensor model by the integer electron lattice, yielding a wrapped-Gaussian density parameterized only by read noise and admitting lattice-sum and Jacobi theta-function representations. The invariant is the theta ratio $R=\vartheta_4(q)/\vartheta_3(q)$ with nome $q$, of which any VPM is a contrast normalization; the existing approximations are low-order truncations of its lattice sums, and the amplitude-domain metric converges to its phase-space counterpart at large exposure. A closed-form inverse for read noise in terms of VPM follows from elliptic integrals. The identification yields a moment method of characterization: conversion gain and read noise are estimated jointly from the modulus of the empirical characteristic function of the raw gray values, without specifying the number law, and with leading-order precision depending on it only through its variance.
\end{abstract*}


\section{Introduction}

Single-electron resolution in image sensors becomes achievable when the input-referred read noise ($\sigma$) is sufficiently small \cite{Ma_2017}, enabling histogram-based methods of characterization that are difficult, impractical, or impossible in conventional noise regimes \cite{Starkey_2016,Gach_2022,Nakamoto_2022,Hendrickson_2024,Krynski_2025}. In particular, Starkey and Fossum \cite{Starkey_2016} introduced a practical method for conversion gain and read noise estimation using the photon counting histogram (PCH), including the definition of a valley-peak modulation (VPM) metric.

In the original amplitude-domain definition, VPM is derived from the classical Poisson--Gaussian model and is functionally dependent on both read noise and the Poisson parameter (quanta exposure). This dependence is visible when plotting VPM versus read noise at multiple exposure levels \cite{Starkey_2016}. Nonetheless, exposure-independent approximations are asymptotically accurate for small read noise. The goal of this note is to identify and analyze the underlying object those approximations are probing.

To that end, Section \ref{sec:amplitude_VPM} introduces the sensor model and reviews the amplitude-domain metric with its exposure-independent approximations under the Poisson law. Section \ref{sec:phase_VPM} constructs the phase-space formulation, in which reduction modulo one electron removes the integer electron number exactly, whatever its distribution, leaving a wrapped Gaussian governed by read noise alone. There the earlier approximations reappear as truncations of lattice sums, the amplitude-domain metric is shown to converge to its phase-space counterpart at large exposure, and read noise is recovered from the metric in closed form through elliptic integrals. Section \ref{sec:VPM_not_unique} shows that any VPM is a purpose-selected normalization of a single invariant ratio, the choice amounting to a decision about where the metric spends its sensitivity. Section \ref{sec:estimation_in_phase_space} develops the construction into a practical method of characterization, estimating conversion gain and read noise jointly from raw gray values without imposing a parametric model on the number law.


\section{Amplitude-domain VPM}
\label{sec:amplitude_VPM}

The standard model for a linear pixel's gray values ($X$, in units of digital numbers $(\mathrm{DN})$) is
\begin{equation}
\label{eq:PG_model}
X=\mathcal M_K+\mathcal M_K^\prime\sigma Z,
\end{equation}
where $\mathcal M_k=\mu+k/g$ is the electron-to-DN transfer function with DC offset $\mu\,(\mathrm{DN})$ and conversion gain $g\,(e\text{-}/\mathrm{DN})$, $\mathcal M_k^\prime\coloneqq\partial_k\mathcal M_k$, $K\sim F_K$ is the number of electrons accumulated over the integration time, with law $F_K$ supported on the nonnegative integers but otherwise unspecified, $Z\sim\operatorname{Gaussian}(0,1)$ is independent of $K$, and $\sigma$ is the read noise $(e\text{-})$. For the linear transfer function, $\mathcal M_k^\prime=1/g$, so the read noise referred to DN is $\sigma/g$. In practice, $X$ is rounded by an analog-to-digital converter, an effect accounted for by decomposing $\sigma$ into analog input-referred noise and quantization noise components as $\sigma^2=\sigma_\mathrm{read}^2+\sigma_\mathrm{quant}^2$, with $\sigma_\mathrm{quant}^2= g^2/12+O((q_\mathrm{read})^{1/g^2})$, $q_\mathrm{read}=e^{-2(\pi\sigma_\mathrm{read})^2}$.

Conditioning on $K$, $X|K\sim\operatorname{Gaussian}(\mathcal M_K,(\mathcal M_K^\prime\sigma)^2)$, and thus we obtain the marginal density
\begin{equation}
f_X(x)=\sum_{k=0}^\infty\mathsf P(K=k)\,\frac{1}{\mathcal M_k^\prime\sigma}\phi\left(\frac{x-\mathcal M_k}{\mathcal M_k^\prime\sigma}\right),\qquad \phi(x)=\frac{1}{\sqrt{2\pi}}e^{-x^2/2}.
\end{equation}
The amplitude-domain theory is constructed under the Poisson law, $K\sim\operatorname{Poisson}(H)$ with $\mathsf P(K=k)=e^{-H}H^k/k!$, where the quanta exposure $H\,(e\text{-})$ is the mean number of accumulated electrons; this assumption is retained for the remainder of the section, rendering \eqref{eq:PG_model} the familiar Poisson--Gaussian model. For $\sigma \lesssim 0.5\,e\text{-}$, $f_X$ has local maxima (peaks) near $x=\mathcal M_k$ and local minima (valleys) near $x=\mathcal M_{k+1/2}$.

In \cite{Starkey_2016}, an amplitude-domain VPM metric was introduced to relate valley filling to read noise. For brevity, we note that this amplitude-domain metric is invariant w.r.t.\ $\mu$ and $g$, so one may set $\mu=0$ and $g=1$, leaving a function of only $H$ and $\sigma$. One representative form used in practice is
\begin{equation}
\label{eq:VPM_x_kappa_def}
\operatorname{VPM}_x
\coloneqq
1-\frac{2f_X(v_\kappa)}{f_X(p_\kappa)+f_X(p_{\kappa+1})},\qquad\kappa\coloneqq\left\lfloor\tfrac{1}{2}\big(\sqrt{4H^2+1}-1\big)\right\rfloor.
\end{equation}
By this definition, $\{\kappa,\kappa+1\}$ are the two most probable electron numbers, $\{p_\kappa,p_{\kappa+1}\}$ are the corresponding DN-referred peak locations of $f_X$, and $v_\kappa$ is the location of the central valley between them. Because $\operatorname{VPM}_x$ is built from extrema of the Poisson--Gaussian density, its domain is limited. As $\sigma$ increases the peaks of $f_X$ merge, and the defining peak-valley pair vanishes at an exposure-dependent cutoff $\sigma^\ast(H)$, beyond which $\operatorname{VPM}_x$ is undefined.

\begin{figure}[htbp]
    \centering
    \includegraphics[width=0.7\linewidth]{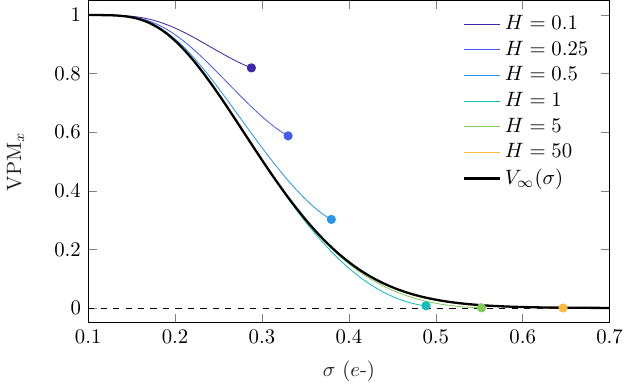}
    \caption{$\operatorname{VPM}_x$ as a function of $\sigma$ for selected $H$-values (cf. \cite{Starkey_2016}, Fig.~4). Filled circles mark the cutoffs $\sigma^\ast(H)$ beyond which the defining peak-valley pair ceases to exist; the black curve $V_\infty$ is identified in Section \ref{subsec:phase_VPM_SF_connect}. Exposures below one electron approach $V_\infty$ from above, integer exposures from below.}
    \label{fig:VPM_calc_v2}
\end{figure}

Figure \ref{fig:VPM_calc_v2} plots a numerical evaluation of $\operatorname{VPM}_x$ for several $H$-values, with filled circles marking the cutoffs $\sigma^\ast(H)$. As $\sigma\to 0^+$, all curves approach unity because overlap between neighboring Gaussian components vanishes exponentially; exposure dependence enters only through an exponentially small correction. As $H$ increases, the curves are observed to approach a single common curve, shown in black and denoted $V_\infty(\sigma)$, whose identity is taken up in Section \ref{subsec:phase_VPM_SF_connect}.

Motivated by these empirical observations, Starkey and Fossum presented three exposure-independent approximations for $\operatorname{VPM}_x$, obtained by locally truncating $f_X$ to a finite sum of Poisson-weighted Gaussian components (valid in the small-$\sigma$ regime) and exploiting a local symmetry of the central Poisson probabilities that becomes increasingly accurate as $H$ increases. Under these assumptions one can derive, in increasing order of accuracy (cf.~\cite{Starkey_2016}, Eqs.~(12)--(14)),
\begin{subequations}
\label{eq:VPMx_aprx}
\begin{equation}
\label{eq:VPMx_1}
\operatorname{VPM}_x\approx1-2e^{-1/(8\sigma^2)}
\end{equation}
\begin{equation}
\label{eq:VPMx_2}
\operatorname{VPM}_x\approx1-\frac{2e^{-1/(8\sigma^2)}}{1+2e^{-1/(2\sigma^2)}}
\end{equation}
\begin{equation}
\label{eq:VPMx_3}
\operatorname{VPM}_x\approx1-\frac{2e^{-1/(8\sigma^2)}+2e^{-9/(8\sigma^2)}}{1+2e^{-1/(2\sigma^2)}}.
\end{equation}
\end{subequations}
We next show that these approximations are precisely low-order lattice-sum truncations of a law-invariant VPM defined in phase space.


\section{Law-invariance via phase-space transformation}
\label{sec:phase_VPM}

To motivate the phase-space formulation we return to the general model \eqref{eq:PG_model} and define the inverse transfer function $\mathcal M_x^{-1}=g(x-\mu)$. Then,
\begin{equation}
\mathcal M_X^{-1}=K+\sigma Z,
\end{equation}
recovering $K$ up to the additive read-noise term. To remove $K$, and with it all dependence on $F_K$, the natural operation is reduction modulo one electron. A convenient smooth representation of ``mod $1$'' is the unit-circle embedding $y\mapsto e^{i2\pi y}$, which identifies real values that differ by integers since $e^{i2\pi (n+y)}=e^{i2\pi y}$ for any $n\in\mathbb Z$.

Using this fact we define
\begin{equation}
\label{eq:phase_xform_exact}
\Phi\coloneqq\operatorname{Arg}\big(e^{i2\pi\mathcal M_X^{-1}}\big)\overset{\mathrm{a.s.}}{=}\operatorname{Arg}\big(e^{i2\pi\sigma Z}\big),
\end{equation}
where $\operatorname{Arg}z\in(-\pi,\pi]$ denotes the principal argument (phase). Since $e^{i2\pi K}=1$ for \emph{any} integer-valued random variable $K$, the almost-sure identification \eqref{eq:phase_xform_exact} makes no reference to $F_K$; it follows that $\Phi\sim\operatorname{WrappedGaussian}(0,(2\pi\sigma)^2)$ for every $F_K$. The Poisson assumption of the amplitude-domain construction thus plays no role in phase space. In the Poisson case the distribution of $\Phi$ is, in particular, independent of $H$, which is precisely the exposure invariance probed by the approximations \eqref{eq:VPMx_aprx}. More generally, any non-ideality that perturbs the electron statistics (e.g., integration-time jitter, temporal variation in illumination intensity or dark-current generation) leaves $f_\Phi$, and with it every phase-space quantity derived from it, unchanged.

\begin{figure}[htbp]
    \centering
    \includegraphics[width=0.7\linewidth]{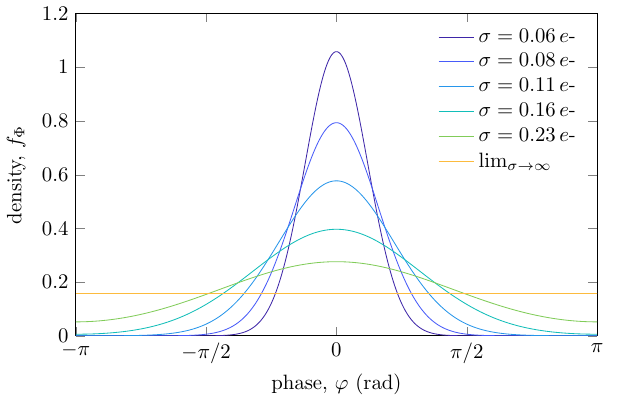}
    \caption{The phase density $f_\Phi$ for several values of $\sigma$, including the uniform limit as $\sigma\to\infty$.}
    \label{fig:WrappedNormal_calc}
\end{figure}

As $\sigma\to0^+$, the density $f_\Phi$ vanishes at the boundaries $\varphi=-\pi,\pi$, indicating the peaks in $f_X$ are nearly resolved. Conversely, as $\sigma\to\infty$, $\Phi\to\mathrm{Uniform}(-\pi,\pi]$, indicating complete loss of peak structure. Figure~\ref{fig:WrappedNormal_calc} illustrates $f_\Phi$ for several values of $\sigma$.


\subsection{Wrapped Gaussian series and Jacobi theta functions}

Since $\Phi=2\pi\sigma Z$ reduced modulo $2\pi$ onto $(-\pi,\pi]$, its density is the sum over wraps,
\begin{equation}
\label{eq:wrap_sum}
f_\Phi(\varphi)
=\frac{1}{2\pi}
\sum_{n=-\infty}^\infty
\frac{1}{\sigma}\phi\left(\frac{n+\varphi/2\pi}{\sigma}\right).
\end{equation}
Poisson summation recasts \eqref{eq:wrap_sum} as
\begin{equation}
f_\Phi(\varphi)=\frac{1}{2\pi}\bigg(1+2\sum_{n=1}^\infty q^{n^2}\cos n\varphi\bigg)=\frac{1}{2\pi}\vartheta_3\!\left(\frac{\varphi}{2},q\right),\quad q=e^{-2(\pi\sigma)^2}
\end{equation}
identifying $f_\Phi$ with the Jacobi $\vartheta_3$ with argument $\varphi/2$ and nome $q$. Technically speaking, $f_\Phi$ is a $2\pi$-periodic function in $\varphi$ on the entire real line; restricting $\varphi\in(-\pi,\pi]$ renders it a probability density. Peak and valley heights are then
\begin{equation}
f_\Phi(0)=\frac{1}{2\pi}\vartheta_3(q),\qquad f_\Phi(\pi)=\frac{1}{2\pi}\vartheta_4(q),
\end{equation}
with $\vartheta_m(q)\coloneqq\vartheta_m(0,q)$ denoting the Jacobi constants (classically, \emph{Thetanullwerte}). Equivalently, using \eqref{eq:wrap_sum},
\begin{equation}
\label{eq:f_Phi_peak_valley_lattice}
f_\Phi(0)\propto1+2\sum_{n=1}^\infty e^{-n^2/(2\sigma^2)},\qquad
f_\Phi(\pi)\propto2\sum_{n=0}^\infty e^{-(2n+1)^2/(8\sigma^2)}.
\end{equation}


\subsection{Phase-space VPM and connection to earlier approximations}
\label{subsec:phase_VPM_SF_connect}

By analogy to the amplitude-domain definition, define the phase-space VPM as
\begin{equation}
\label{eq:phase_VPM_def}
\operatorname{VPM}_\varphi
\coloneqq
1-\frac{f_\Phi(\pi)}{f_\Phi(0)}
\end{equation}
and let
\begin{equation}
m(q)\coloneqq(\vartheta_2(q)/\vartheta_3(q))^4,\quad\text{satisfying}\quad \left(\vartheta_4(q)/\vartheta_3(q)\right)^4=1-m(q)
\end{equation}
denote the inverse elliptic nome. Then
\begin{equation}
\label{eq:VPM_inv_nome}
\operatorname{VPM}_\varphi=1-\sqrt[4]{1-m\big(e^{-2(\pi\sigma)^2}\big)}.
\end{equation}
Alternatively, using the lattice-sum specializations \eqref{eq:f_Phi_peak_valley_lattice},
\begin{equation*}
\operatorname{VPM}_\varphi=\underset{\substack{n_1\to\infty\\ n_2\to\infty}}{\lim}v(n_1,n_2),\qquad
v(n_1,n_2)=
1-\frac{2\sum_{n=0}^{n_1}e^{-(2n+1)^2/(8\sigma^2)}}{1+2\sum_{n=1}^{n_2}e^{-n^2/(2\sigma^2)}}.
\end{equation*}
It follows that $v(0,0)$, $v(0,1)$, and $v(1,1)$ reproduce the three exposure-independent approximations given in \cite{Starkey_2016} (Eqs.~\eqref{eq:VPMx_1}--\eqref{eq:VPMx_3}), respectively.

The correspondence extends beyond the approximations. Returning to Figure \ref{fig:VPM_calc_v2}, the black curve $V_\infty$ is a plot of $\operatorname{VPM}_\varphi$ itself; the approach of the curves to $V_\infty$ thus displays the collapse of $\operatorname{VPM}_x$ onto $\operatorname{VPM}_\varphi$ with increasing $H$. The limit can be established directly. Adopting for this argument the normalization $\mu=0$, $g=1$ of Section \ref{sec:amplitude_VPM} and writing $\pi_k(H)$ for the Poisson probabilities, center and scale $f_X$ about the $\kappa$ of Eq.~\eqref{eq:VPM_x_kappa_def},
\begin{equation*}
\mathcal A(\varphi,H)\coloneqq\frac{f_X(\kappa+\varphi/2\pi)}{2\pi\,\pi_\kappa}=\frac{1}{2\pi}\sum_{n=-\kappa}^{\infty}\frac{\pi_{\kappa+n}}{\pi_\kappa}\,\frac{1}{\sigma}\phi\left(\frac{\varphi/2\pi-n}{\sigma}\right),\quad\varphi\in\mathbb R
\end{equation*}
in terms of which
\begin{equation*}
\operatorname{VPM}_x=1-\frac{2\mathcal A\big(2\pi(v_\kappa-\kappa),H\big)}{\mathcal A\big(2\pi(p_\kappa-\kappa),H\big)+\mathcal A\big(2\pi(p_{\kappa+1}-\kappa),H\big)}.
\end{equation*}
For fixed $n$, the ratios satisfy $\pi_{\kappa+n}/\pi_\kappa=1+O(1/H)$ so the Poisson probabilities become locally uniform about their mode as $H\to\infty$. Unimodality and the definition of $\kappa$ bound every ratio by $\max(1,\pi_{\kappa+1}/\pi_\kappa)\leq\sqrt2$, so dominated convergence applies and
\begin{equation*}
\mathcal A(\varphi,H)\to f_\Phi(\varphi),\quad H\to\infty.
\end{equation*}
Since the ratios are independent of $\varphi$, the same domination applied to the series of suprema over any compact $\varphi$-interval makes the convergence locally uniform, and likewise for the $\varphi$-derivatives. Consequently, the limiting periodic density has nondegenerate extrema precisely at the integer multiples of $\pi$ and the central extrema of $f_X$ converge to them, with $p_\kappa-\kappa\to0$, $v_\kappa-\kappa\to\tfrac12$, and $p_{\kappa+1}-\kappa\to1$. Combining the three limits and noting that $f_\Phi(2\pi)=f_\Phi(0)$ gives
\begin{equation*}
\lim_{H\to\infty}\operatorname{VPM}_x=1-\frac{2f_\Phi(\pi)}{f_\Phi(0)+f_\Phi(2\pi)}=\operatorname{VPM}_\varphi.
\end{equation*}
The large-$H$ limit thus simultaneously removes the lower summation boundary, flattens the local Poisson weights, and places the central extrema on the lattice, so that the three sources of exposure dependence in $\operatorname{VPM}_x$, each a feature of sampling $f_X$ at extrema of the Poisson-weighted density, vanish together. This convergence is one instance of a broader correspondence in which functionals of $f_X$ converge to their phase-domain counterparts at large $H$; a rigorous treatment of this correspondence, including rates, will be presented elsewhere. The limit object, meanwhile, requires no limit at all. For every $F_K$ and every $\sigma>0$,
\begin{equation*}
\operatorname{VPM}_\varphi=1-\frac{\sum_{n\in\mathbb Z}f_X(\mathcal M_{n+1/2})}{\sum_{n\in\mathbb Z}f_X(\mathcal M_{n})}
=1-\frac{\big(\sum_{k=0}^\infty\mathsf P(K=k)\big)\sum_{n^\prime\in\mathbb Z}\phi\big(\frac{n^\prime+1/2}{\sigma}\big)}{\big(\sum_{k=0}^\infty\mathsf P(K=k)\big)\sum_{n^\prime\in\mathbb Z}\phi\big(\frac{n^\prime}{\sigma}\big)},
\end{equation*}
the lattice sites being fixed by the transfer function alone and the weights summing to unity across the full lattice, leaving a function independent of $F_K$ entirely.

A further useful representation, for the purpose of numerical evaluation, follows from the Jacobi triple-product identities \cite[\S20.5(i)]{NIST:DLMF}
\begin{equation*}
\operatorname{VPM}_\varphi=\lim_{n_1\to\infty}v^\ast(n_1),\qquad v^\ast(n_1)=1-\prod_{n=0}^{n_1}\tanh^2\big((2n+1)(\pi\sigma)^2\big),
\end{equation*}
with the truncated product expressible in closed form as a ratio of $q$-Pochhammer symbols. Since each $\tanh^2(\cdot)$ factor is a sigmoid in $\sigma$, $v^\ast(n_1)$ is monotone decreasing and $[0,1]$-valued with the correct limits $1$ and $0$, unlike the truncations $v(n_1,n_2)$.

\begin{figure}[htbp]
    \centering
    \includegraphics[width=0.7\linewidth]{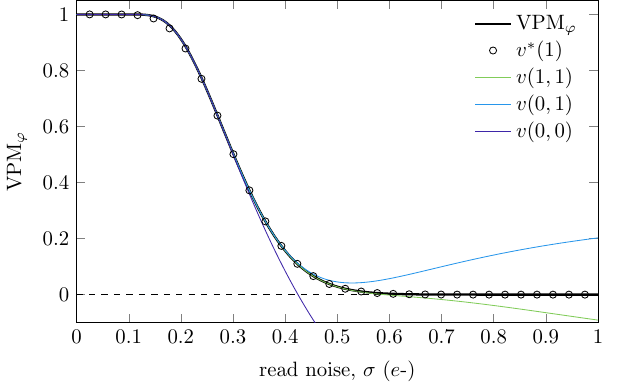}
    \caption{Phase-space VPM and the three truncation-based approximations reported in \cite{Starkey_2016}, together with the truncated product $v^\ast(1)$.}
    \label{fig:VPM_phi_calc}
\end{figure}

Figure \ref{fig:VPM_phi_calc} plots $\operatorname{VPM}_\varphi$ with its various approximations. The aptness of the DSERN regime definition ($\sigma\leq 0.5\,e\text{-}$) is apparent, as this is the threshold where the VPM markedly departs from zero. As read noise decreases further, VPM approaches unity rapidly. Evaluating \eqref{eq:VPM_inv_nome} gives $\operatorname{VPM}_\varphi(0.15\,e\text{-})\approx 0.99$, marking the transition into an ultra-low-noise regime corresponding to nearly zero peak overlap in the amplitude domain \cite{Ma_2015_2} where the quanta counting accuracy of the pixel is for all practical purposes unity \cite{Hendrickson_PCA}.


\subsection{Inverting the phase-space VPM}

Read noise can also be recovered explicitly from the phase-space VPM. Rearranging \eqref{eq:VPM_inv_nome} gives
\begin{equation}
m(q)=1-(1-\operatorname{VPM}_\varphi)^4,
\qquad q=e^{-2(\pi\sigma)^2}.
\end{equation}
Equivalently,
\begin{equation}
\label{eq:inv_VPM_qForm}
e^{-2(\pi\sigma)^2}=q(m),\qquad m=1-(1-\operatorname{VPM}_\varphi)^4,
\end{equation}
where $q(m)$ is the elliptic nome
\begin{equation}
q(m)\coloneqq\exp\left(-\pi\frac{\mathcal K(1-m)}{\mathcal K(m)}\right),\quad \text{with}\quad \mathcal K(m)\coloneqq\int_0^{\pi/2}\frac{dt}{\sqrt{1-m\sin^2 t}}
\end{equation}
denoting the complete elliptic integral of the first kind. Isolating $\sigma$ then yields the desired inverse
\begin{equation}
\label{eq:VPM_inv}
\sigma=
\frac{1}{\sqrt{2\pi}}\bigg(
\frac{\mathcal K((1-\operatorname{VPM}_\varphi)^4)}
{\mathcal K(1-(1-\operatorname{VPM}_\varphi)^4)}\bigg)^{1/2}.
\end{equation}
For numerical work the elliptic integrals need not be evaluated directly. By the Legendre product representation of the arithmetic--geometric mean, $\operatorname{agm}(1,b)$ \cite{WolframAGMProduct},
\begin{equation*}
\sigma=\lim_{n_1\to\infty}s(n_1),\quad s(n_1)=\frac{1}{\sqrt{2\pi}}\prod_{n=0}^{n_1}\bigg(\frac{1+k_n}{1+\ell_n}\bigg)^{1/2},
\end{equation*}
with $k_0=(1-\operatorname{VPM}_\varphi)^2$, $\ell_0=\big(1-(1-\operatorname{VPM}_\varphi)^4\big)^{1/2}$, $k_{n+1}=2\sqrt{k_n}/(1+k_n)$, and $\ell_{n+1}=2\sqrt{\ell_n}/(1+\ell_n)$. The iteration is quadratically convergent, the number of correct digits roughly doubling per factor, so that a handful of factors yield the limit to machine precision.


\section{Interpretation and non-uniqueness of VPM}
\label{sec:VPM_not_unique}

The phase-space formulation identifies the law-invariant theta ratio
\begin{equation}
R\coloneqq\frac{f_\Phi(\pi)}{f_\Phi(0)}=\frac{\vartheta_4(q)}{\vartheta_3(q)}
\end{equation}
as the fundamental object governing valley-peak modulation. In particular, the results derived in Section \ref{subsec:phase_VPM_SF_connect} follow directly from truncations of the lattice-sum representation of $R$ and the definition $\operatorname{VPM}_\varphi=1-R$. However, any mapping
\begin{equation}
V = \Psi(R),
\end{equation}
with $\Psi:[0,1]\to[0,1]$ continuous, strictly decreasing, and satisfying $\Psi(0)=1$ and $\Psi(1)=0$, defines an equally valid VPM metric.

The structure of $R$ does not privilege a canonical normalization. Instead, these conditions fix the total variation of every such metric at unity, $\int_0^\infty \big|\partial_\sigma V\big|\,\mathrm{d}\sigma = 1$, so that the total resolving power is a fixed unit budget distributed over the noise axis. Reparametrization by $\Psi$ redistributes this budget; thus, selecting a normalization is a decision about where a metric spends its sensitivity, dictated by intended use. The choice $\Psi(R)=1-R$ concentrates its budget in the DSERN regime and saturates in the ultra-low-noise regime $(\sigma\leq0.15\,e\text{-})$, where \eqref{eq:f_Phi_peak_valley_lattice} gives $1-\operatorname{VPM}_\varphi\sim2e^{-1/(8\sigma^2)}$. This saturation is itself informative. The quanta counting error probability of an idealized uniform quantizer obeys the same exponential law, $\operatorname{erfc}\big(1/(2\sqrt{2}\sigma)\big)\sim (\sqrt{2/\pi}\sigma)2e^{-1/(8\sigma^2)}$ \cite{Hendrickson_PCA}, so in this regime $\operatorname{VPM}_\varphi$ functions as a quanta-number-resolving accuracy indicator, its plateau reporting that counting accuracy has effectively saturated (cf.\ Section~\ref{subsec:phase_VPM_SF_connect}). The Michelson contrast variant
\begin{equation*}
V\coloneqq\frac{1-R}{1+R}= \sqrt[4]{m(q^4)},
\qquad
\sigma = \frac{1}{\sqrt{8\pi}}\left(\frac{\mathcal K(1-V^4)}{\mathcal K(V^4)}\right)^{1/2},
\end{equation*}
is an equally valid normalization with a closely related allocation. Another choice is the peak-to-mean contrast $V=1-\sqrt{\operatorname{agm}(1,R^2)}$, comparing the peak of $f_\Phi$ to the uniform level $1/(2\pi)$. Here, $1-V\sim\sqrt{2\pi}\sigma$ as $\sigma\to0^+$ so that, unlike the saturating normalizations above, this metric is asymptotically linear in $\sigma$ and allocates its unit budget nearly uniformly at small read noise. Accordingly, the fundamental invariant is $R$; a specific VPM arises only after choosing a purpose-dependent normalization of it.


\section{Conversion gain and read noise estimation in phase space}
\label{sec:estimation_in_phase_space}


\subsection{Characteristic-function factorization and gain locking}

To develop the phase-space construction into a method of characterization, we consider \eqref{eq:PG_model} with ADC rounding, $X=\operatorname{round}\big(\mu+(K+\sigma_\mathrm{read}Z)/g\big)$. Up to aliasing terms of order $(q_\mathrm{read})^{1/g^2}$ (Section~\ref{sec:amplitude_VPM}), the modulus of the population characteristic function $c(\omega)\coloneqq\mathsf E\,e^{i2\pi\omega X}$ factors as
\begin{equation}
\label{eq:rounded_factor}
|c(\omega)|=\big|\varphi_K(2\pi\omega/g)\big|\,q_\mathrm{read}^{(\omega/g)^2}\,\lvert\operatorname{sinc}\pi\omega\rvert,
\end{equation}
where $\varphi_K$ is the characteristic function of $F_K$ and $\operatorname{sinc}x=\sin(x)/x$. Since $\log|\operatorname{sinc}\pi\omega|=-(\pi\omega)^2/6-(\pi\omega)^4/180-\cdots$, the quadratic content of the quantization envelope is absorbed exactly by the replacement $\sigma_\mathrm{read}^2\to\sigma^2=\sigma_\mathrm{read}^2+g^2/12$, recovering the decomposition of Section~\ref{sec:amplitude_VPM}, and the rounded model reduces to
\begin{equation}
\label{eq:mag_factor}
|c(\omega)|=\big|\varphi_K(2\pi\omega/g)\big|\,q^{(\omega/g)^2},\quad q=e^{-2(\pi\sigma)^2}
\end{equation}
up to the quartic and higher $\operatorname{sinc}$ terms. The read noise accessible to estimation is therefore the total $\sigma$.

Provided $F_K$ is unit-span, i.e., the gcd of the differences between points in its support is one, $|\varphi_K(2\pi\omega/g)|\le1$ with equality precisely at $\omega\in g\mathbb Z$. The first factor of \eqref{eq:mag_factor} therefore forms a periodic comb whose maxima lie exactly on the gain lattice, with $|c(ng)|=q^{n^2}$ irrespective of $\mu$. The second factor, a Gaussian envelope, displaces the maximizer of $|c(\omega)|$ near $g$ to $\approx g(1-\rho)$, where $\rho=\sigma^2/(\mathsf{Var}K+\sigma^2)$ (cf.~\cite{Hendrickson_SPIE_2023}, Eqs.~(14) and (18)). The gain is therefore estimated by locking onto this local maximum and iteratively dividing out the envelope. A search bracket about the nominal gain (here $\pm15\%$) isolates the fundamental mode, excluding the trivial maximum at $\omega=0$ and the harmonics near $\omega=ng$, $n\ge2$. Pass~$0$ locks the raw modulus, $g_0=\arg\max_\omega|c(\omega)|$ and sets $q_0=|c(g_0)|$. Pass $k+1$ then relocks on the flattened objective,
\begin{equation}
\label{eq:lock_iter}
g_{k+1}\leftarrow\argmax_{\omega}\,|c(\omega)|\,q_k^{-(\omega/g_k)^2},
\qquad
q_{k+1}\leftarrow|c(g_{k+1})|.
\end{equation}
At the fixed point the flattening is exact and the corrected objective reduces to $|\varphi_K(2\pi\omega/g)|$, whose fundamental maximum occurs exactly at $g$. Hence $(g,q)$ is the fixed point of \eqref{eq:lock_iter} under the unit-span condition, requiring no knowledge of $\mu$ or of $F_K$; the only external input is the coarse search interval about the nominal gain.


\subsection{Precision and calibration sensitivity}

In practice the iteration runs on the empirical characteristic function of a random sample of gray values $\mathbf x=\{x_1,\dots,x_N\}$,
\begin{equation}
\label{eq:circ_moments}
\hat c(\omega)=\frac1N\sum_{j=1}^N e^{i2\pi\omega x_j},
\end{equation}
in place of $c$, producing iterates $(\hat g_k,\hat q_k)$, with the terminal estimates denoted $(\hat g,\hat q)$. Evaluating $\hat c$ at integer multiples of the estimated gain, $\hat c_n=\hat c(n\hat g)$, estimates the circular moments of the phase variable up to an offset phase carrying $\mu$, which the modulus removes. By the delta method, the estimates obey, to first order,
\begin{equation}
\label{eq:lock_precision}
\mathsf{Var}(\hat g/g)\approx\frac{(1-q^4)}{8\pi^2 q^2N\,\mathsf{Var}K},\quad
\mathsf{Var}\,\hat\sigma\approx\frac{(1-q^2)^2}{32\pi^4\sigma^2 q^2N},\quad
\mathsf{Cov}(\hat g,\hat\sigma)\approx\frac{g(1-q^2)^2}{16\pi^4\sigma q^2N\,\mathsf{Var}K};
\end{equation}
a detailed analysis of the iteration will be presented in future work. The division of labor between algorithm and law is as follows: the iteration and its fixed point require no model of $F_K$; the leading-order precision \eqref{eq:lock_precision} depends on $F_K$ only through $\mathsf{Var}K$; higher-order corrections to the bias and convergence rate involve higher cumulants of $K$. Every phase-space quantity then follows in closed form,
\begin{equation}
\label{eq:VPM_moment_est}
\hat q=|\hat c(\hat g)|,\qquad
\widehat{\operatorname{VPM}}_\varphi=1-\frac{\vartheta_4(\hat q)}{\vartheta_3(\hat q)},\qquad
\hat\sigma=\sqrt{\frac{-\log\hat q}{2\pi^2}},
\end{equation}
and inverting $\widehat{\operatorname{VPM}}_\varphi$ through \eqref{eq:VPM_inv} returns the same $\hat\sigma$, the forward and inverse phase-space maps being exact. The higher moduli provide a built-in consistency check of the wrapped-Gaussian model through the population prediction $|c(ng)|=q^{n^2}$ tested empirically by $|\hat c_n|\approx\hat q^{\,n^2}$.

Under \eqref{eq:mag_factor}, the sensitivity of $\hat\sigma$ to residual calibration error is governed by an exact law. Writing $\hat g=g(1+\varepsilon)$, the population modulus \eqref{eq:mag_factor} evaluated at $\hat g$ factors into the read-noise term and $\lvert\varphi_K(2\pi\varepsilon)\rvert$; since $e^{i2\pi K}=1$ for integer $K$, only the detuning survives. For any $K$ with finite fourth moment, the estimator therefore obeys the exact population-level law
\begin{equation}
\label{eq:gain_error_general}
\hat\sigma^{\,2}=(1+\varepsilon)^2\sigma^2-\frac{\log\lvert\varphi_K(2\pi\varepsilon)\rvert}{2\pi^{2}}=(1+\varepsilon)^2\sigma^2+\varepsilon^{2}\mathsf{Var}K+O(\varepsilon^{4}),
\end{equation}
the expansion following from the cumulant series of $\log\varphi_K$, of which the modulus retains only the even-order terms; odd cumulants of $K$ drop out alongside the offset $\mu$. The distributional invariance noted after \eqref{eq:phase_xform_exact} is thus exact at $\varepsilon=0$, with the law of $K$ re-entering only through imperfect gain knowledge; to second order, this dependence is entirely through $\mathsf{Var}K$. Under the Poisson model of Section \ref{sec:amplitude_VPM}, every cumulant equals $H$ and \eqref{eq:gain_error_general} reduces to (cf.\ \cite{Hendrickson_2023}, Eq.~(23))
\begin{equation}
\label{eq:gain_error_law}
\hat\sigma^{\,2}=(1+\varepsilon)^2\sigma^2+H\varepsilon^2\operatorname{sinc}^2\pi\varepsilon=(1+\varepsilon)^2\sigma^2+H\varepsilon^2+O(\varepsilon^4).
\end{equation}
For the locked calibration, $\varepsilon$ is zero-mean with variance given by $\mathsf{Var}(\hat g/g)$ in \eqref{eq:lock_precision}, feeding back through \eqref{eq:gain_error_law} at $O(1/N)$ and thus negligibly at all practical sample sizes; for an externally supplied calibration, a percent-scale $\varepsilon$ dominates the error budget of $\hat\sigma$ once $N\gtrsim2\times10^{4}$.


\subsection{Monte Carlo validation}

\begin{figure}[htb]
    \centering
    \includegraphics[width=\linewidth]{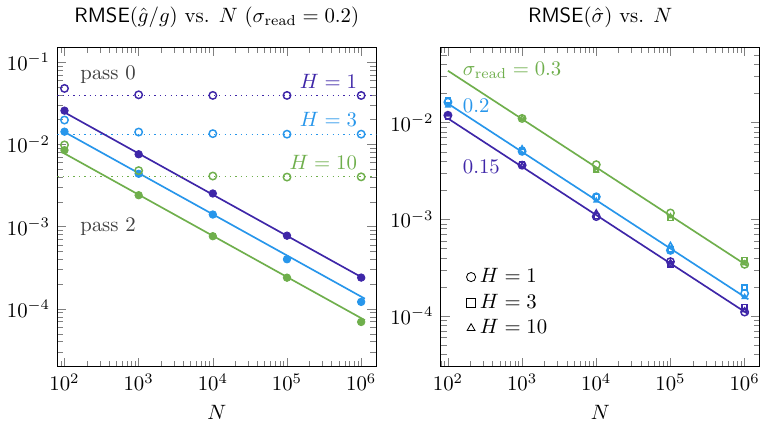}
    \caption{Monte Carlo behavior of \eqref{eq:circ_moments}--\eqref{eq:lock_precision} under the rounded data model of this section ($60$--$400$ trials per point; operating points with appreciable lock-failure rates, $q\sqrt{N}\lesssim4$, omitted). Left: relative RMSE of $\hat g$ at $\sigma_\mathrm{read}=0.2$. Open symbols mark the raw (pass-$0$) lock, which stalls on its bias floor $\rho$ (dotted); filled symbols mark pass $2$, which follows the first law of \eqref{eq:lock_precision} (solid curves) with no fitted parameters. Right: RMSE of $\hat\sigma$ against the second law of \eqref{eq:lock_precision}. Symbol shapes distinguish $H\in\{1,3,10\}$; their collapse onto common curves in the right panel reflects the absence of $\mathsf{Var}K$ from that law.}
    \label{fig:gain_lock}
\end{figure}

Using this model, we constructed a Monte Carlo experiment for the complete procedure with $K\sim\operatorname{Poisson}(H)$, $(g,\mu)=(0.0645\,e\text{-}/\mathrm{DN},\,150\,\mathrm{DN})$, sweeping $H\in\{1,3,10\}$, $\sigma_\mathrm{read}\in\{0.15,0.2,0.3\}$, and $N\in\{10^2,\dots,10^6\}$ at $60$--$400$ trials per operating point. Before running the procedure, we verified that the neglected $|\operatorname{sinc}\pi\omega|$ tail displaces the fixed point by less than $2\times10^{-6}\,e\text{-}$ in read noise at every operating point, far below sampling noise. Figure~\ref{fig:gain_lock} presents the results of the procedure. The raw lock $\hat g_0$ stalls on its bias floor $\rho$; two corrected passes remove the bias at every operating point with $N\ge10^3$, leaving estimates whose standard deviations follow \eqref{eq:lock_precision}; since the simulation rounds just as an ADC does while the iteration divides out only the Gaussian envelope, the agreement also confirms the negligibility of the tail neglected in \eqref{eq:mag_factor}. The pair \eqref{eq:lock_iter} and \eqref{eq:VPM_moment_est} thus constitutes a characterization method in its own right, a moment-type procedure delivering $(\hat g,\hat\sigma)$, and with them every admissible VPM, from a single sample of gray values with no knowledge of $\mu$ or the law of $K$, and hence no model of $K$ to misspecify. The price of this universality is efficiency. The procedure consumes only the modulus of the empirical characteristic function near the gain lattice, deliberately discarding the information a parametric model of $K$ would supply, so when such a model is trusted, as with the Poisson model of Section \ref{sec:amplitude_VPM}, the maximum-likelihood methods of \cite{Hendrickson_2023,Hendrickson_2024} characterize all model parameters at lower variance, with \eqref{eq:lock_iter} supplying their initialization. What the discard purchases is robustness rather than optimality. Under misspecification of the electron-number model the estimates remain valid, their precision \eqref{eq:lock_precision} depending on $F_K$ only through $\mathsf{Var}K$, converting model risk into statistical variance.


\section{Discussion and conclusion}

Amplitude-domain VPM is influenced by both read noise and quanta exposure. The exposure dependence is a consequence of sampling the Poisson--Gaussian density at its extrema, whose existence and location are set by the local Poisson weights; summation over the full electron lattice removes both this dependence and the domain restriction, and the phase mapping realizes this summation exactly. From the perspective of this note, VPM is fundamentally a circular modulation metric: once the integer electron lattice is quotiented out, the native space of the metric is the unit circle rather than the real-valued amplitude domain. The resulting wrapped Gaussian depends only on the read noise parameter $\sigma$, yielding a phase-space VPM invariant to the electron-number law, with a closed-form theta ratio and an explicit inverse mapping. Practically, this invariance permits direct comparison of read-noise estimates across differing exposures and illumination statistics, and the closed-form inversion \eqref{eq:VPM_inv} is the exact population-level relationship between the metric and $\sigma$, removing the truncation-selection choice implicit in the original approximations.

The contribution of this note is thus both structural and, following from the structure, methodological. Structurally, the Fourier-analytic treatment of the sensor model does more than reorganize known results; it exposes an invariant of the model itself. The conventional Poisson--Gaussian formulation foregrounds the electron-number law, whereas the invariant belongs to the additive integer-lattice structure shared by every such law; the wrapped Gaussian, the nome, and the theta ratio $R$ are its exact expression. Methodologically, once the invariant is visible, characterization under it follows. The moment method of Section~\ref{sec:estimation_in_phase_space} estimates conversion gain and read noise jointly from raw gray values, trading statistical efficiency, relative to likelihood methods that assume the Poisson law \cite{Hendrickson_2023,Hendrickson_2024}, for validity under every unit-span $F_K$, with precision and calibration-error laws depending, at leading order, on $F_K$ only through its variance. The metric itself retains its role as a quotable, law-invariant summary of read noise whose saturation carries direct quanta-number-resolving meaning. The present analysis resolves the base case of a linear transfer function and thereby opens the corresponding question for nonlinear transfer, where the electron levels are no longer equally spaced. Such structural clarifications form part of a larger effort aimed at placing image sensor characterization on an explicit mathematical footing.


\begin{backmatter}
\bmsection{Funding}
No funding was received for this work.

\bmsection{Disclosures}
The authors declare no conflicts of interest.

\bmsection{Data availability}
The simulated data and code underlying the results presented in this paper are not publicly available at this time but may be obtained from the authors upon reasonable request.
\end{backmatter}

\bibliography{sources}

\end{document}